%% file: master.tex
\documentclass{article}
\usepackage{xcolor}

\usepackage{type1cm}        
\usepackage{makeidx}         
\usepackage{graphicx}        
\usepackage{multicol}        
\usepackage[bottom]{footmisc}

\usepackage{newtxtext}       %
\usepackage{newtxmath}       

\usepackage{amsmath}
\usepackage{amssymb}
\usepackage{amsfonts}
\usepackage{enumerate}
\usepackage{algorithm}
\usepackage{algorithmic}
\usepackage{MnSymbol}

\makeindex             

\newcommand{\figref}[1]{Figure \ref{#1}}

\newcommand{\secref}[1]{Section~\ref{#1}}

\newcommand{\algoref}[1]{Algorithm \ref{#1}}
\newcommand{\etal}{et~al.}
\newcommand{\x}{\mathbf{x}}
\newcommand{\y}{\mathbf{y}}
\newcommand{\n}{\mathbf{n}}
\newcommand{\m}{\mathbf{m}}
\newcommand{\tn}{\widetilde{\n}}
\newcommand{\q}{\mathbf{q}}
\newcommand{\bv}{\mathbf{v}}
\newcommand{\bc}{\mathbf{c}}
\newcommand{\bh}{\mathbf{h}}
\newcommand{\f}{\mathbf{f}}
\newcommand{\g}{\mathbf{g}}
\newcommand{\M}{\mathbb{M}}
\newcommand{\R}{\mathbb{R}}
\newcommand{\Z}{\mathbb{Z}}
\newcommand{\V}{\mathcal{V}}
\newcommand{\Rn}{\mathcal{R}}
\newcommand{\B}{\mathcal{B}}
\newcommand{\D}{\mathcal{D}}
\newcommand{\RG}{\mathcal{RG}}
\newcommand{\RS}{\mathbb{W}}
\newcommand{\JCN}{\mathrm{JCN}}
\newcommand{\MRS}{\mathrm{MRS}}
\newcommand{\MDRG}{\mathrm{MDRG}}
\newcommand{\MPAIR}{\mathrm{MPAIR}}
\newcommand{\Q}{\mathbf{Q}}
\newcommand{\s}{\varphi}
\newcommand{\simt}{\widetilde{\varphi}}
\newcommand{\Sim}{\Phi}
\newcommand{\Simt}{\overline{\Phi}}

\begin{document}

\title{A Topological Similarity Measure between Multi-Field Data using Multi-Resolution Reeb Spaces}

\author{Tripti Agarwal \thanks{ International Institute of Information Technology, Bangalore, India. {\tt tripti.agarwal@iiitb.org}}
\and 
Yashwanth Ramamurthi \thanks{ International Institute of Information Technology, Bangalore, India. {\tt yashwanth@iiitb.org}}  
\and
Amit Chattopadhyay \thanks{International Institute of Information Technology, Bangalore, India. {\tt
            a.chattopadhyay@iiitb.ac.in}}
}
%
%
\date{}
\maketitle

\begin{abstract}
Searching topological similarity between a pair of shapes or data is an important problem in data analysis and visualization.
The problem of computing similarity measures using scalar topology has been studied extensively and proven useful in shape and data matching. Even though multi-field (or multivariate) topology-based techniques reveal richer topological features, research on computing  similarity measures using multi-field topology is still in its infancy. In the current paper, we propose a novel  similarity measure between two piecewise-linear multi-fields
based on their multi-resolution Reeb spaces - a newly developed data-structure that captures the topology of a multi-field. Overall, our method consists of two steps: (i) building a  multi-resolution Reeb space corresponding to each of the multi-fields and (ii) proposing a similarity measure for a list of matching pairs (of nodes), obtained by comparing the  multi-resolution Reeb spaces. 
We demonstrate an application of the proposed similarity measure by detecting the nuclear scission point in a time-varying multi-field data from computational physics.
\end{abstract}


\input{body.tex}

\end{document}

%% file: body.tex


\section{Introduction} 
Similarity measures using scalar topology have demonstrated significant applications in shape matching, classification of  bio-molecular or protein structures, symmetry detection, periodicity analysis in time-dependant flows, and so on \cite{hilaga2001topology, Zhang2004FastMO, thomas2014multiscale, saikia2014extended}. The design of the similarity algorithms are mostly based on scalar topological data-structures, viz. contour tree, Reeb graph, merge tree, Morse-Smale complex, extremum graph etc. 

Since multi-field topology is richer than scalar topology, using multi-field topology one is expected to design more precise similarity measures for better classification of shapes and data. However, this requires generalizations of the existing data-structures for scalar topology to capture multi-field topology. Towards this, in the current paper, we contribute  as  follows:
\begin{itemize}\itemsep=0.01cm
    \item We introduce a novel Multi-resolution Reeb Space (MRS) data-structure to capture the multi-field topology by generalizing the Multi-resolution Reeb Graph (MRG)  by Hilaga \etal~\cite{hilaga2001topology}.
    
    \item Then, we propose a similarity measure between two MRS structures by generalizing the similarity measure for scalar fields by Hilaga \etal~and Zhang~\etal \cite{hilaga2001topology, Zhang2004FastMO}.
    
    \item Finally, we show the effectiveness of our method by detecting the nuclear scission point
    from our similarity plots for the time-varying  multi-field Fermium-$256$ atom dataset as in ~\cite{2012-Duke-VisWeek}.
\end{itemize}

The next section discusses the related works on topological similarity measures. \secref{sec:background} provides the necessary background to understand our method. \secref{sec:MultiResolutionReebSpace} and \secref{sec:similarity} describe our algorithms for computing an MRS and computing the proposed similarity measure between two MRSs, respectively. Finally, \secref{sec:application} shows an application of our method and concludes with a summary.


\section{Related Work}
\label{sec:related}
Topological similarity and distance measures between scalar fields have been studied extensively.  
Beketayev \etal \cite{beketayev2014measuring} propose an interleaving distance as a distance  between merge trees.
Bauer \etal \cite{bauer2014measuring} propose a stable functional distortion metric for computing distance between two Reeb graphs. 
Saikia et al. \cite{saikia2014extended} develop an extended branch decomposition graph (eBDG)  and identify the repeating topological structure in a scalar data. 
Thomas \etal \cite{thomas2014multiscale} propose a multiscale symmetry detection technique using contour clustering.  
Other work has been done to find similarity between scalar fields by proposing a distance metric between merge trees \cite{beketayev2014measuring}. 
Saikia \etal \cite{saikia2015fast} propose a histogram feature descriptor to differentiate between subtrees of a merge tree. 
Narayanan \etal \cite{narayanan2015distance} present a distance measure to compare scalar fields using extremum graphs. Sridharamurthy \etal \cite{2020-Sridharamurthy} present an edit distance based method between merge trees for feature visualization in time-varying scalar field data.
Among the multi-resolution techniques, the most important ones are by Hilaga \etal \cite{hilaga2001topology} - a similarity measure based on multi-resolution Reeb graphs (MRG) and by Zhang \etal \cite{Zhang2004FastMO} - topology matching using multi-resolution dual contour trees.
The current paper generalizes these techniques for multi-fields. 

However, research on computing topological similarity measures for multi-fields is still at a nascent stage. Recently, Agarwal \etal \cite{agarwal2019topological} have proposed a distance metric between two multi-fields based on their fiber-component distributions of  and demonstrate its usefulness over scalar-topology. It is worth mentioning few important data-structures for capturing multi-field topology. Carr \etal \cite{JCN_paper} develop a joint contour net (JCN) for a quantized approximation of the Reeb space. Duke \etal \cite{2012-Duke-VisWeek} successfully apply the JCN to visualize nuclear scission features in multi-field density data.
 Chattopadhyay \etal \cite{2014-EuroVis-short, 2015-Chattopadhyay-CGTA-simplification}  propose a hierarchical multi-dimensional Reeb graph (MDRG) structure equivalent to the Reeb space.  These data-structures are useful for the development of the current algorithm.

\section{Background}
\label{sec:background}
In this section, we describe the necessary background to understand the proposed similarity measure between MRSs. 
More precisely, we briefly highlight the important tools for capturing the scalar and multi-field topology. 

\paragraph{Piecewise-Linear Multi-Fields.}
 Most of the data in scientific visualization comes as a discrete set of real values  at every vertex (grid-point) of a mesh in a volumetric domain. 
Let us consider the data domain as a compact $m$-dimensional manifold $\mathcal{M}$ and let $\M$ be a triangulation (mesh)  of $\mathcal{M}$  whose vertices contain the data values. Let $\mathbf{V}(\M)=\{\bv_0, \bv_1, \ldots, \bv_p\}$ be the set of vertices of $\M$. An $n$-dimensional multi-field data can be described by a vertex map $\hat{\f}=(\hat{f_1},\hat{f_2},\ldots,\hat{f_n}):\mathbf{V}(\mathbb{M})\rightarrow \mathbb{R}^n$ which maps each vertex to a $n$-tuple of scalar values. From this discrete map we define a  piecewise-linear (PL) multi-field $\f=(f_1,f_1,\ldots,f_n):\mathbb{M}\rightarrow \mathbb{R}^n$ as
$\f(\x)=\sum_{i=0}^p\alpha_i \hat{\f}(\bv_i)$
where $\x \in \sigma$ (a simplex of $\M$) has a unique convex combination of its vertices that can be expressed as $\x=\sum_{i=0}^p\alpha_i \bv_i$ with $\alpha_i\geq 0$ and $\sum_{i=0}^p \alpha_i=1$. We note, $\f$ is continuous and the restriction of $\f$ over each simplex of $\M$ is linear. In the current paper, we consider the PL multi-field $\f:\M\rightarrow \R^n$ with $m\geq n\geq 1$ as our input multi-field.
In particular, if $n=1$, $f:\M\rightarrow\mathbb{R}$ is a PL scalar field.

\paragraph{Reeb Space and Reeb Graph.}
Given a PL multi-field $\f:\M\rightarrow\R^n$ and a range value $\bc\in \R^n$, the inverse image $\f^{-1}(\bc)=\{\x\in \M: \f(x)=\bc \}$ is called a \emph{fiber} and a connected component of the fiber is called a \emph{fiber-component} ~\cite{2004-Saeki,Saeki2014}. Each fiber-component is an equivalence class obtained by an equivalence relation $\sim$ on $\M$: $\x \sim \y\Leftrightarrow \x,\, \y\in \M,\; \f(\x)=\f(\y)=\bh$ and $\x,\, \y$ belong to the same connected component of $\f^{-1}(\bh)$.  This equivalence relation partitions $\M$ into the set of all equivalence classes or fiber-components. The space $\RS_\f$ formed by the set of equivalence classes along with the topology induced by a quotient map $q_\f: \M\rightarrow \RS_\f$  is called the Reeb space \cite{2008-edels-reebspace}. The quotient map $q_\f$ maps each point of $\M$ to its equivalence class. Geometrically, under some regularity conditions, $\RS_\f$ is an $n$-dimensional polyhedron. 


In particular, for  a PL scalar field $f:\M\rightarrow\R$,  the fiber $f^{-1}(c)=\{x\in \M: f(x)=c \}$ is known as a \textit{level set} of the isovalue $c\in \R$ and a connected component of the level set is called a \textit{contour} instead of fiber-component. Under some regularity conditions, the Reeb space of the scalar field $f:\M\rightarrow\R$ is a $1$-dimensional CW-complex or a graph structure, known as the Reeb graph and is denoted by $\RG_f$. Therefore, $\RG_f$ consists of a set of \emph{nodes} and \emph{arcs},  each arc connecting two of the nodes. Each point of the Reeb graph corresponds to a contour. In particular, the nodes of the Reeb graph correspond to the contours passing through the critical points \cite{2003-Cole-McLaughlin-Loop-Reeb} of $f$ and the arcs connecting the nodes represent the contours which pass through the regular points (not critical!) of $f$. 

\begin{figure}[h!]
        \centering        \includegraphics[width=\textwidth]{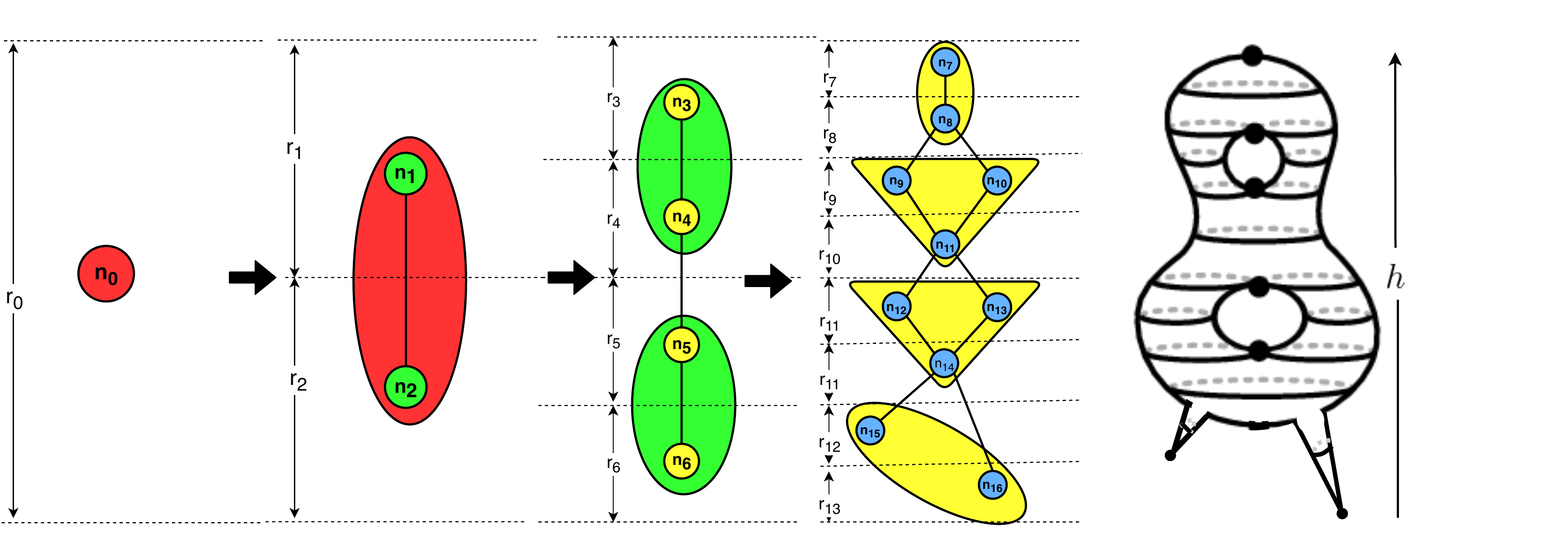}
        \caption{An MRG of the height function $h$ of a standing double torus with legs. The figure shows the MRG with four Reeb graphs at four different resolutions - coarser to finer Reeb graphs are shown from the left to right.}
        \label{fig:MRG}
\end{figure}

\paragraph{Multi-Resolution Reeb Graph.}
A multi-resolution Reeb graph of a PL scalar field $h:\M\rightarrow \R$, proposed by Hilaga \etal~\cite{hilaga2001topology}, is a data-structure that computes a finite series of Reeb graphs at various levels of data resolutions. In practice, each Reeb graph is obtained by subdividing the data range into a set of $Q=2^k,\; (k=0, 1,\ldots,N-1)$ levels of resolution by a dyadic subdivision (as shown in \figref{fig:MRG}). The domain $\M$ is partitioned into fat (or quantized) contours accordingly and then the Reeb graph at that resolution is obtained by constructing the adjacency graph of the fat contours~\cite{hilaga2001topology}. The Reeb graphs in an MRG satisfy the following properties: 1. a parent-child relationship is maintained between the nodes of the adjacent Reeb graphs at consecutive levels, 2. by repeating the process of subdivision, when the levels of resolution goes to infinity the MRG converges to the actual Reeb graph of $h$ and 3. a Reeb graph at a particular resolution contains all the information about the coarser resolution Reeb graphs. \figref{fig:MRG} shows the MRG of the height field of a standing double torus with legs, using $4$ resolutions. 

\begin{figure}[!h]
        \centering
        \includegraphics[width=\textwidth]{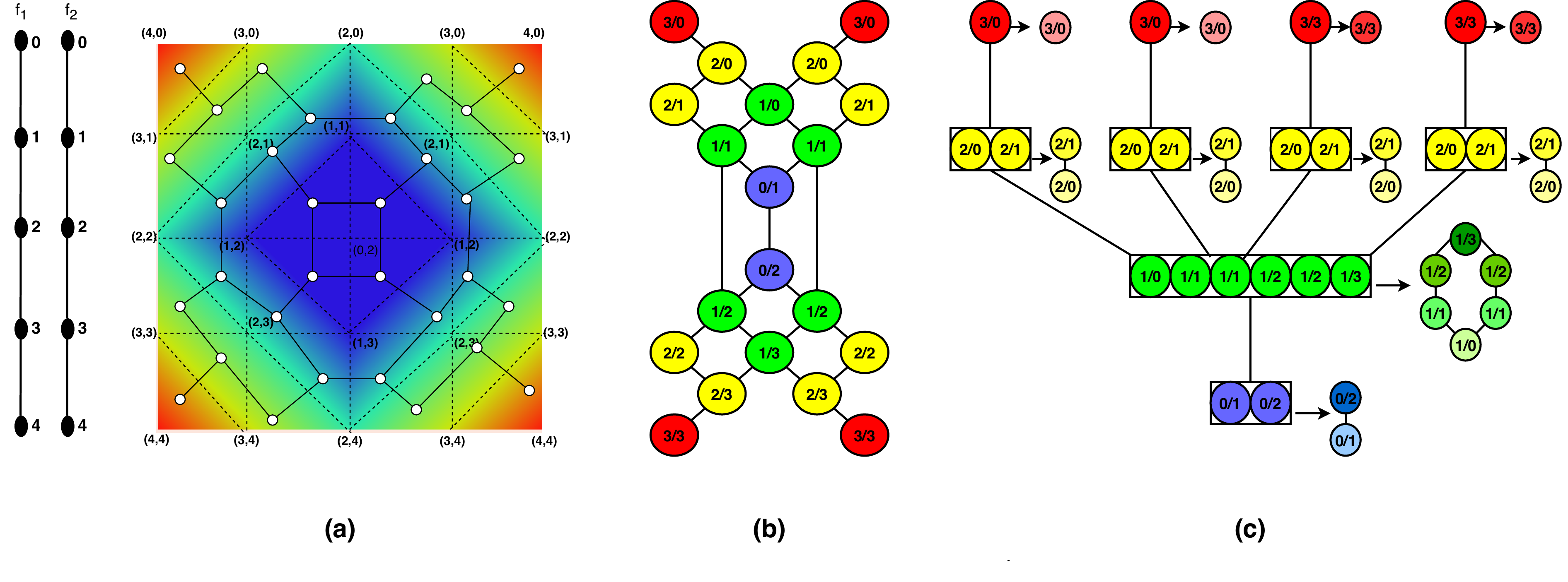}
        \caption{(a) A PL bivariate data over a 2D mesh, (b) JCN at $4\times 4$ resolution, (c) MDRG computed using the algorithm in \cite{2015-Chattopadhyay-CGTA-simplification}.}
        \label{fig:JCN_Mdrg}
\end{figure}

\paragraph{Joint Contour Net.} A Joint Contour Net (JCN) of a PL multi-field $\f=(f_1,\ldots,f_n):\M\rightarrow\R^n$, proposed by Carr \etal~\cite{JCN_paper}, is a quantized approximation of the Reeb space using a chosen number of quantization levels (or levels of resolution). 
The data range is first quantized or subdivided into $Q=q_1\times q_2\times \ldots\times q_n$ levels of quantization (here, range of $f_i$ is quantized into $q_i$ levels, $i=1,2, \ldots,n$), similar as MRG. Thus the range of $\f$ is discretized into a finite set, such as a subset of $\Z^n$.
In this case, for a quantized range value $\bh\in\Z^n$, instead of a fiber we obtain a \emph{quantized fiber} or \emph{joint level set}, denoted as $\tilde{\f}^{-1}(\bh)=\{\x\in \M: round(\f(\x))=\bh \}$ where the $round$ function is applied on each component. Quantized fibers are not always connected. A connected component of 
a quantized fiber is called a \emph{quantized fiber-component} or \emph{joint contour}. The joint contour net is an adjacency graph of the joint contours or quantized fiber-components.
Figure~\ref{fig:JCN_Mdrg}(a) shows an example of JCN corresponding to a simulated bivariate dataset. In our method, to compute a multi-resolution Reeb space corresponding to a multi-field, we compute joint contour nets at different resolutions.


\paragraph{Multi Dimensional Reeb Graph.}
    A Multi Dimensional Reeb Graph (MDRG) of a PL multi-field $f:\M\rightarrow\R^n$, proposed by Chattopadhyay \etal \cite{2014-EuroVis-short, 2015-Chattopadhyay-CGTA-simplification}, is a hierarchical decomposition of the Reeb space (or the corresponding joint contour net) into a set of Reeb graphs in different dimensions. In particular, to construct the MDRG for a PL bivariate field $\f=(f_1, f_2):\M\rightarrow\R^2$, we first compute the Reeb graph $\RG_{f_1}$ of the field $f_1$ (in the first dimension). Now each point  $p\in \RG_{f_1}$ corresponds to a contour $C_p$ of $f_1$. We restrict function $f_2$ (second dimension) on $C_p$ and define the restricted function $\widetilde{f}_2^p=f_2|_{C_p}$. Then for the second dimension, we compute Reeb graphs $\RG_{\widetilde{f}_2^p}$ for each of these restricted functions $\widetilde{f}_2^p$. 
    Thus MDRG of $f$, denoted by $\MDRG_f$, can be defined as: $\MDRG_f =\{(p_1, p_2): p_1\in \RG_{f_1}, p_2\in \RG_{\widetilde{f}_2^{p_1}}\}$. 
    The definition can be extended for any PL multi-field $\f:\M\rightarrow\R^n$ with $m\geq n>2$. Fig.~\ref{fig:JCN_Mdrg}(b) shows an example of a quantized Reeb space or JCN for a PL bivariate field (in Fig.~\ref{fig:JCN_Mdrg}(a)) and Fig.~\ref{fig:JCN_Mdrg}(c) shows its MDRG. 
    
\section{Multi-Resolution Extension of Reeb Space}
\label{sec:MultiResolutionReebSpace}
In this section, we develop a new multi-resolution Reeb space structure that captures the topology of a PL multi-field data at different resolutions, similar as MRG of a PL scalar field~\cite{hilaga2001topology}. The Reeb space at a particular resolution is approximated by the JCN. The idea is to develop a series of JCNs at various resolutions.

\subsection{Overview}
\begin{figure}[h!]
        \centering
        \includegraphics[width=\textwidth]{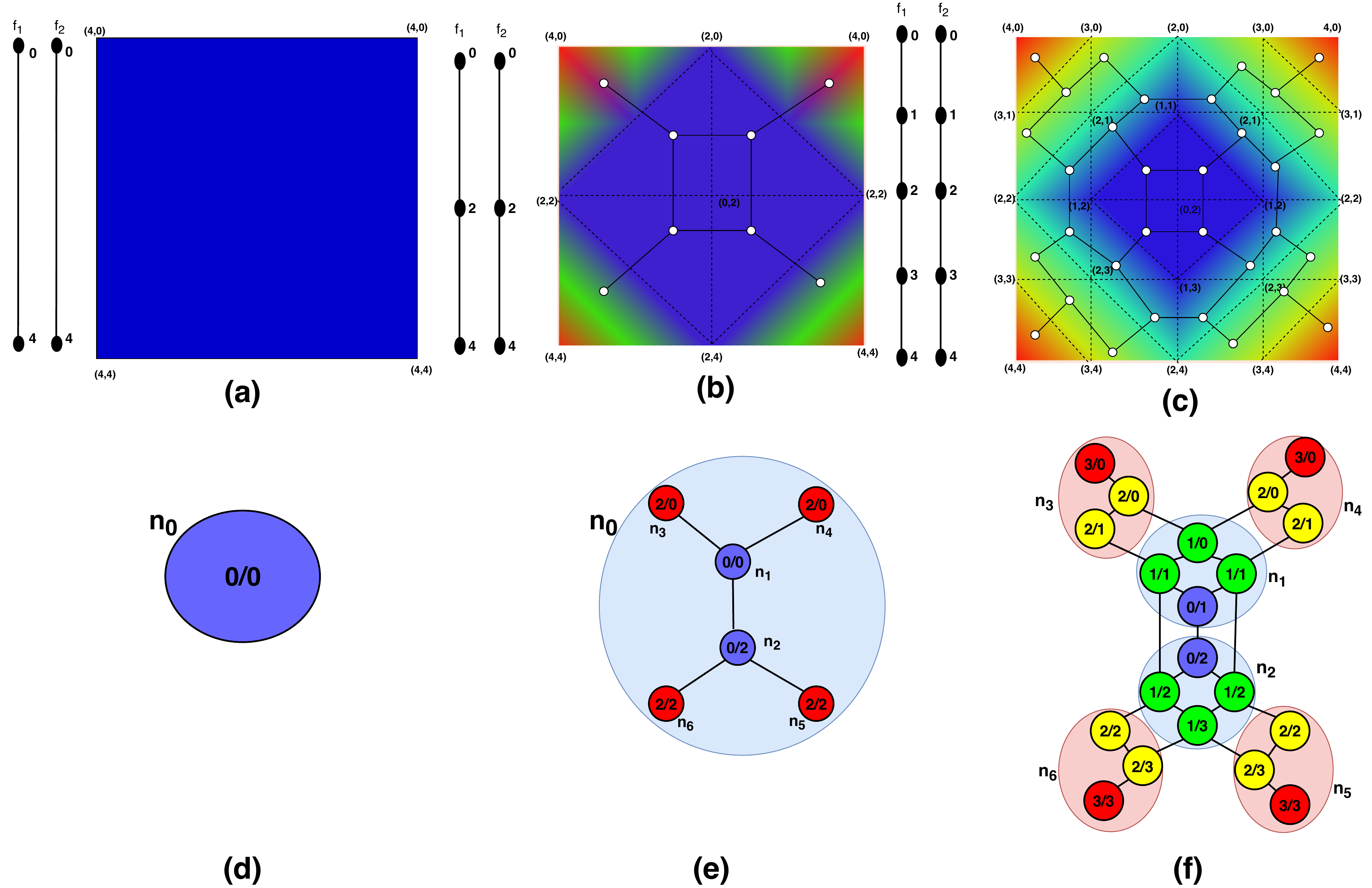}
        \caption{Multi-resolution Reeb Space corresponding to a PL bivariate data: (ring, height). (a) Each component field is quantized into one level, and the corresponding JCN in (d) consists of only one node $n_0$. (b) Each component field is quantized into two levels and the corresponding JCN is shown is (e). (c) Each component field is quantized into four levels, and the corresponding JCN is shown in (f). Parent-child relationships between the nodes of JCNs in consecutive resolutions are shown, e.g. $n_0$ is parent of $\{n_1,n_2,n_3,n_4,n_5,n_6\}$.}
        \label{fig:MRS}
\end{figure}

Let $\f=(f_1,f_2,\ldots,f_n):\M\rightarrow\R^n$ be a PL multi-field and $\JCN_\f(\q)$ denote the JCN of $\f$ with parameter-vector $\q=(q_1, q_2, \ldots,q_n)$, each parameter $q_i$ being the levels of resolution of the component field $f_i$ for $i\in \{1, 2, \ldots, n\}$. Thus, using $\q$, we obtain an approximated Reeb space $\JCN_\f(\q)$ with $Q=q_1\times q_2\times\ldots\times q_n$ levels of resolution of the multi-field $\f$. Now to construct a multi-resolution Reeb space, for simplicity, we consider a finite sequence   $\{\q^{(k)}=(2^k, 2^k, \ldots, 2^k):k=0, 1, \ldots, N-1\}$ of $N\, (\geq 1)$ increasing levels of resolution of $\f$. Corresponding to this sequence of resolutions, we obtain a sequence $\left\{\JCN_\f(\q^{(k)}):k=0, 1, \ldots, N-1\right\}$ of $N$ approximated Reeb spaces that defines a \emph{Multi-resolution Reeb Space} (MRS) of $\f$ with $N$ levels of resolution and is denoted by $\MRS_{\f,\,N}$. Moreover, $\JCN_\f(\q^{(k)})$ or $\JCN_{\f,\,k}$ (in short) for $k=0, 1,\ldots, N-1$ of the MRS satisfy the following properties:

\begin{enumerate}[(P1)]
    \item There are parent-child relationships between the nodes of adjacent resolutions, i.e between the nodes in $\JCN_{\f,\,k}$ and $\JCN_{\f,\,k+1}$. For example, in \figref{fig:MRS}, node $n_0$ of the JCN in (d) is the parent of the nodes $\{n_1, n_2, n_3, n_4, n_5, n_6\}$ of the JCN in (e).
    
    \item Note that when the levels of resolution goes to infinity the JCN graph converges to the Reeb space of $\f$, i.e. $\JCN_{\f,\,k}$ converges to the Reeb space  $\RS_\f$ as $k\rightarrow\infty$ (see \cite{2015-Chattopadhyay-CGTA-simplification} for details). In otherwords, we can say the multi-resolution Reeb space $\MRS_{\f,\,N}$ converges to $\RS_\f$ as the levels of resolution $N$ tends to $\infty$.
    
    \item A JCN of a certain resolution in $\MRS_{\f,\,N}$ implicitly contains all the information of the JCNs of coarser resolutions, i.e. $\JCN_{\f,\,k}$ contains all the information of $\JCN_{\f,\,0},\,\JCN_{\f,\,1},\,\ldots, \JCN_{\f,\,k-1}$. Once a JCN of a certain resolution is constructed, the coarser resolution JCN can be constructed by grouping the adjacent nodes in same coarser range interval, as shown in \figref{fig:MRS}.
\end{enumerate}

\subsection{Construction of the Multi-Resolution Reeb Space}
Consider the parameter vector  $\q^{(N-1)}=(2^{N-1}, 2^{N-1}, \ldots, 2^{N-1})$ of (dyadic) levels of resolutions corresponding to PL multi-field $\f=(f_1, f_2,\ldots,f_n)$ where $N\,(\geq 1)$ is an integer.
 Then the JCN of $\f$, using $\q^{(N-1)}$, can be constructed using the algorithm described by Carr \etal~\cite{JCN_paper} and is denoted by $\JCN_\f(\q^{(N-1)})$ or $\JCN_{\f,\,N-1}$. To construct the multi-resolution Reeb space $\MRS_{\f,N}$, we start with $\JCN_{\f,\,N-1}$ and construct its  coarser resolution Reeb spaces, sequentially, by merging each pair of consecutive range intervals and grouping the adjacent nodes in the corresponding intervals. This is  similar to the construction of MRG by Hilaga \etal \cite{hilaga2001topology}. \algoref{algo:coarserreeb} outlines the method for constructing a coarser resolution Reeb space $\JCN_{\f,\,k-1}$
from the finer resolution Reeb space $\JCN_{\f,\,k}$. Let $R_{\f}=[f_1^{\min},\,f_1^{\max}]\times [f_2^{\min},\,f_2^{\max}]\times\ldots\times [f_n^{\min},\,f_n^{\max}]$ be the $n$-dimensional range interval of the multi-field $\f$ where $f_i^{\min}=\displaystyle\min_{\x\in \M} f_i(\x)$ and $f_i^{\max}=\displaystyle\max_{\x\in \M} f_i(\x)$ for $i=1, 2, \ldots, n$.  The range  $[f_i^{\min},\,f_i^{\max}]$ of the component field $f_i$ (for $i= 1, 2, \ldots, n$) is subdivided into $q=2^{k}$ ($k\in \mathbb{N}:$ set of natural numbers) dyadic levels of resolution or sub-intervals: 
$r_0^{(i)}=[x_0^{(i)},\,x_1^{(i)}),\,r_1^{(i)}=[x_1^{(i)},\,x_2^{(i)}),\ldots, r_{q-1}^{(i)}=[x_{q-1}^{(i)},\,x_q^{(i)}]$, where $x_0^{(i)}=f_i^{\min}$ and $x_q^{(i)}=f_i^{\max}$ ($i=1, 2, \ldots, n$). Thus the range $R_{\f}$ is subdivided into $Q=q\times q\times \ldots \times q $  ($n$ times) dyadic levels of resolution or sub-intervals, denoted by $r_{i_1i_2\ldots i_n}=r_{i_1}^{(1)}\times r_{i_2}^{(2)}\times \ldots \times r_{i_n}^{(n)}$ (where $i_1, i_2, \ldots, i_n=0, 1, \ldots, q-1$). For computing the coarser level Reeb space $\JCN_{\f,\,k-1}$ (where $k\geq 1$) we merge the adjacent levels of $\JCN_{\f,\,k}$ (in pairs) and obtain the coarser sub-intervals as $R_{i_1 i_2\ldots i_n}=\mathrm{merge}(r_{2i_1}^{(1)},r_{2i_1+1}^{(1)} )\times \mathrm{merge}(r_{2i_2}^{(2)},r_{2i_2+1}^{(2)} )\times \ldots\times \mathrm{merge}(r_{2i_n}^{(n)},r_{2i_n+1}^{(n)})$ (for $i_1, i_2, \ldots, i_n=0, 1, \ldots, \frac{q}{2}-1$). So the levels of resolution of $\JCN_{\f,\,k-1}$ reduces to $\tilde{Q}=\frac{q}{2}\times\frac{q}{2}\times\ldots\times\frac{q}{2}$. We construct a Union-Find structure UF \cite{1975-Tarjan-UF} from the adjacency of the nodes of $\JCN_{\f,\,k}$ with ranges in   $R_{i_1 i_2\ldots i_n}$. Each connected component of UF becomes a node of the coarser Reeb space $\JCN_{\f,\,k-1}$ and the adjacencies of these new nodes are determined by the adjacencies of the components in  $\JCN_{\f,\,k}$.

\begin{algorithm}[H]
\caption{\sc{CreateCoarserReebSpace}}
\label{algo:coarserreeb}
\textbf{Input:} $\JCN_{\f,\,k}$\\
\textbf{Output:} $\JCN_{\f,\,k-1}$
\begin{algorithmic}[1]
\FOR{$i_1=0$ to $\frac{q}{2}-1$}
\FOR{$i_2=0$ to $\frac{q}{2}-1$}
\STATE $\ddots$
\FOR{$i_n=0$ to $\frac{q}{2}-1$}
\STATE \% RANGE MERGING FOR COARSER RESOLUTION
\STATE $R_{i_1 i_2\ldots i_n}=\mathrm{merge}(r_{2i_1}^{(1)},r_{2i_1+1}^{(1)} )\times \mathrm{merge}(r_{2i_2}^{(2)},r_{2i_2+1}^{(2)} )\times \ldots\times \mathrm{merge}(r_{2i_n}^{(n)},r_{2i_n+1}^{(n)} ) $
\STATE Create Union-Find Structure UF for the nodes of $\JCN_{\f, k}$ in range $R_{i_1 i_2\ldots i_n}$ 
\STATE \% CREATING NODES OF COARSER JCN
\FOR{ each component $C_j$ in UF}
\STATE Create a node $n_{C_j}$ in $\JCN_{\f,\,k-1}$
\STATE Map node-ids and field-values of the finer JCN
\STATE Set node $n_{C_j}$ as the parent for the nodes in $C_j$
\ENDFOR

\ENDFOR
\ENDFOR
\STATE $\udots$
\ENDFOR 
\STATE \% ADDING EDGES IN COARSER JCN
\FOR{ each edge $e_1e_2$ in $\JCN_{\f,\,k}$}
\IF {$e_1, e_2 \in$ components $C_j \neq C_l$ and $\f(e_1) \neq \f(e_2)$}
\STATE Add edge $e(n_{C_j}, n_{C_l})$ in $\JCN_{\f,\,k-1}$ if not already present
\ENDIF
\ENDFOR
\RETURN{$\JCN_{\f,\,k-1}$}
\end{algorithmic}
\end{algorithm}

\figref{fig:MRS} illustrates an MRS for a simple bivariate  data in a 2D box domain with $3$ levels of resolution. The construction of the MRS starts with the construction of the Reeb space with $Q=2^2\times 2^2$ levels at the finest resolution. 

\subsection{Node Attributes for Similarity Measure}
To obtain a quantitative similarity measure between two multi-resolution Reeb spaces, we associate several attributes to the nodes of the MRS that quantify different topological and geometrical properties of the multi-field data, similar as in ~\cite{hilaga2001topology, Zhang2004FastMO}. The attribute set corresponding to a node $\n$, denoted by $\tn$, is defined as $\tn=\{\V(\n), \Rn(\n), \B_0(\n), \D(\n)\}$ where
$\V(\n)=\frac{\text{Volume}(\n)}{\text{Total Volume} (\M)}$ is the normalized volume of the node $\n$, 
$\Rn(\n)=\frac{\text{measure}(\text{range}(\n))}{\text{measure}(\text{range}(\f))}$ is the normalized range,
$\B_0(\n)$ is the number of components of the joint level set corresponding to $\n$ and
$\D(\n)$ is the degree of $\n$ in the corresponding JCN. 

\section{A Similarity Measure between MRSs}
\label{sec:similarity}
Let $\MRS_{\f,\,N}$ and $\MRS_{\g,\,N}$ 
be two multi-resolution Reeb spaces with same levels of resolution (here, $N$) corresponding to two PL multi-fields $\f$ and $\g$, respectively. Our method of computing the similarity measure between two MRSs has two  steps:
1. Creating a list of matching pairs   from the nodes of the respective MRSs and
2. Computing the similarity measure between the MRSs by defining a similarity between the nodes of a matched pair.
We describe these steps in the following subsections.
\begin{figure}[!h]
    \centering
    \includegraphics[width=\textwidth]{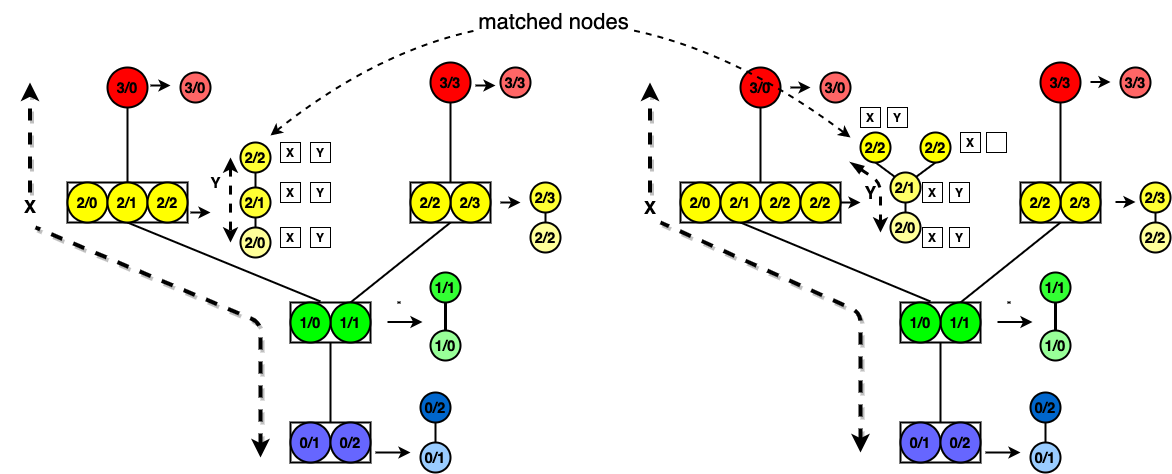}
    \caption{Label propagation is demonstrated for a matching pair in two  MDRGs.  For a bivariate field, two lists of labels need to be maintained. }
    \label{fig:Labelpropagation}
\end{figure}
\subsection{Creating Matching Pairs}
To create the list of matching pairs between two multi-resolution Reeb spaces, denoted by $\mathrm{MPAIR}$, we search from the coarser to the finer resolution Reeb spaces. Nodes $\m\in \MRS_{\f,\,N}$ and $\n\in \MRS_{\g,\,N}$ form a matching pair $(\m,\n)\in \mathrm{MPAIR}$ if they satisfy the following \emph{matching rules} (generalizing the rules in \cite{hilaga2001topology, Zhang2004FastMO}):

\begin{enumerate}[(i)]\itemsep=0.01cm
    \item $\m\in \MRS_{\f,\,N}$ and $\n\in \MRS_{\g,\,N}$ do not belong to any other matched pair,
    
    \item $\m$ and $\n$ belong to the Reeb space of same resolution and both have the same range level,
    
    \item Parent $p(\m)$ of $\m$ and parent $p(\n)$ of $\n$ must have been matched, i.e. $(p(\m),p(\n))$ is already a matching pair, except for the nodes in the coarsest resolution,
    
    \item $\m$ and $\n$ must be topologically consistent. Unlike computing consistency using Reeb graphs in \cite{hilaga2001topology}, here we consider the  MDRGs of the corresponding Reeb spaces. That is, $\m$ and $\n$ should be in the same branches of the respective MDRGs as their matched siblings in all dimensions. 
    In particular, for a bivariate field, to satisfy the topological consistency in each dimension, two lists  of labels are maintained corresponding to each node. Once two nodes are matched and labeled, their siblings in the same branch of the MDRG get the same labels (in each dimension),
    as shown in \figref{fig:Labelpropagation}. 
\end{enumerate}

\noindent
Creating the list $\mathrm{MPAIR}$ of matching pairs between two MRSs is outlined in \algoref{algo:MatchingAlgorithm}.

\begin{algorithm}[H]
\caption{CreatingMatchingPairs}
\label{algo:MatchingAlgorithm}
\textbf{Input:} Multi-resolution Reeb spaces $\MRS_{\f,\,N},\,\MRS_{\g,\,N}$\\
\textbf{Output:} $\mathrm{MPAIR}$ - list of matched pairs
\begin{algorithmic}[1]

\FOR{$ k= 0,1, \ldots, N-1$}

\STATE Add all the nodes of $\JCN_{\f,\, k}$ to a priority queue $\Q$ where the priority of a node is set as its volume attribute.

\WHILE{$\Q$ is not empty}
\STATE $\m\leftarrow \mathrm{Pop}(\Q)$. Search for its best matching pair $\n\in \JCN_{\g,\, k}$, satisfying the \emph{matching rules} (i)-(iv).

\IF{$\n$ is found}  
\STATE $\mathrm{MPAIR}\leftarrow \mathrm{Add}(\{(\m,\n), k\})$
\ENDIF
\ENDWHILE
\ENDFOR
\RETURN $\mathrm{MPAIR}$
\end{algorithmic}
\end{algorithm}

\subsection{Similarity Calculation}
Following Zhang \etal \cite{Zhang2004FastMO}, first we define a real-valued similarity function $\s$ for each matched pair $(\m,\n)\in \MPAIR$
as: 
\begin{align}
\s(\m,\n)=\omega_1 \simt(\V(\m),\V(\n))+\omega_2 \simt(\Rn(\m),\Rn(\n))+\\\nonumber
\omega_3 \simt(\B_0(\m),\B_0(\n))+ \omega_4 \simt(\D(\m),\D_0(\n)) 
\end{align}
where the weights $\omega_i$ satisfy $0\leq \omega_i\leq 1$ for $i=1, 2, 3, 4$ and $\sum_{i=1}^4\omega_i=1$ and the function $\simt: \mathbb{R}^+\times \mathbb{R}^+\rightarrow \mathbb{R}^+$ is defined as $\simt(r_1,r_2)=\frac{\min(r_1,r_2)}{\max(r_1,r_2)} $.
\noindent
Thus, we have $0\leq \s(\m,\n) \leq \s(\m,\m)=\s(\n,\n)=1$.


Next, we define the similarity function $\Sim$ between two $k$-th resolution JCNs $\JCN_{\f,k}$ and $\JCN_{\g,k}$, by the weighed sum of the similarities for all pairs $(\m_i,\n_i)\in \MPAIR$ with $\m_i\in \JCN_{\f,k}$ and $\n_i\in \JCN_{\g,k}$, as
\begin{align}
\Sim(\JCN_{\f,k},\JCN_{\g,k})=\sum_{i=1}^s \frac{\V(\m_i)+\V(\n_i)}{2}\s(\m_i,\n_i)    
\end{align}
where $s$ is the total number of such pairs. Finally, we define the similarity $\Simt$ between two MRSs $\MRS_{\f,N}$ and $\MRS_{\g,N}$ as
\begin{align}
\Simt(\MRS_{\f,N},\MRS_{\g,N})=\frac{1}{N}\sum_{k=0}^{N-1} \Sim(\JCN_{\f,k},\JCN_{\g,k})
\end{align}
We note, $0\leq \Simt(\MRS_{\f,N},\MRS_{\g,N})\leq \Simt(\MRS_{\f,N},\MRS_{\f,N}) = 1$.

\section{Implementation and Application}
\label{sec:application}
\begin{figure}[!h]
    \centering
    \includegraphics[width=\textwidth]{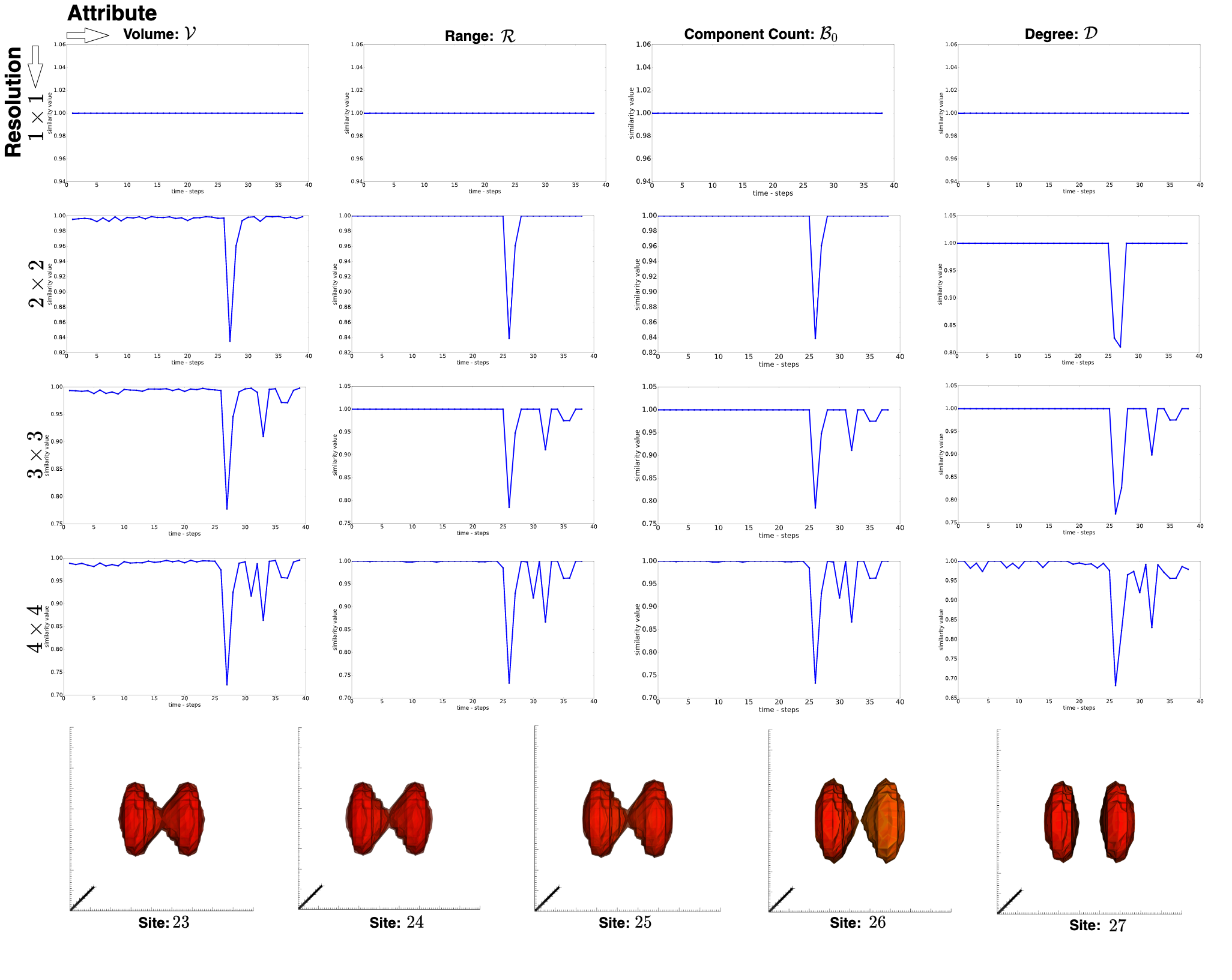}
    \vspace*{-0.2cm}
    \caption{\textbf{Top-four rows:}  Similarity plots for time-varying Fermium atom data. Each row shows the metric plots using multi-resolution Reeb spaces with $4$ different attributes. Each column shows the metric plots using multi-resolution Reeb spaces of $4$ different resolutions.  \textbf{Bottom-row:} Nucleus is visualized at Sites: $23-27$, the split happens at Site $26$ - corresponding to the `lowest' dip in the plots.}
    \label{fig:fermium-results}
\end{figure}

We implement our algorithm for computing  the similarity between two  multi-resolution Reeb spaces  
 under the JCN implementation framework \cite{JCN_paper}. 
As an application of our tool, we consider the time-varying Fermium-$256$ atom dataset as described by Duke \etal~\cite{2012-Duke-VisWeek}. The dataset  is defined on a $19\times19\times19$ sized grid and consists of \emph{proton} and \emph{neutron} densities for $40$ regularly spaced time-steps. \figref{fig:fermium-results} shows the similarity plots by pairwise comparison of the datasets at consecutive time-stamps - for $4$ different resolutions and $4$ attributes. 
From the plots, we see a major topological event at site $26$ which is the \emph{nuclear scission}, as described by Duke \etal \cite{2012-Duke-VisWeek}. 

\section{Conclusion}
In this article, we propose a novel Reeb space based method for measuring the topological similarity between two multi-field data. To compute the similarity measure, we develop a  multi-resolution Reeb space data-structure which converges to the actual Reeb space as the levels of resolution goes to infinity. We show effectiveness of our method in the application of detecting nuclear scission point in a time-varying multi-field data.\\


\bibliographystyle{abbrv}
\bibliography{master.bbl}